# Model Checking Contractual Protocols

Aspassia Daskalopulu
*Department of Computer Science*
*King's College London*
*The Strand, London WC2R 2LS, U.K.*
E-mail: *aspassia@dcs.kcl.ac.uk*

**Abstract**. This paper discusses how model checking, a technique used for the verification of behavioural requirements of dynamic systems, can be usefully deployed for the verification of contracts. A process view of agreements between parties is taken, whereby a contract is modelled as it evolves over time in terms of actions or more generally events that effect changes in its state. Modelling is done with Petri Nets in the spirit of other research work on the representation of trade procedures. The paper illustrates all the phases of the verification technique through an example and argues that the approach is useful particularly in the context of pre-contractual negotiation and contract drafting. The work reported here is part of a broader project on the development of logic-based tools for the analysis and representation of legal contracts [4].

## 1. Introduction

The set of norms which parties agree to abide by during their business exchange essentially specifies the parties' ideal mode of exchange. Some researchers, such as Lee and his associates refer to such norms as "trade procedures" (cf. [9, 1] or "business protocols" (cf. [13]) and they are similar, in nature and function, to what software and hardware engineers call "specifications". Questions of completeness and consistency of such sets of norms arise in the context of contractual negotiation and drafting just as questions of verification arise in relation to specifications in software and hardware engineering. Whether a contract (or trade procedure, business protocol, specification) covers all intended cases without conflicts and whether the ideal mode of exchange that it describes has the appropriate safety ("nothing bad will happen") and liveness ("something good will happen") properties is a concern both during pre-contractual exchanges and when drafters formulate detailed provisions to record the result of such pre-contractual exchanges. Ill-defined contracts may result in undesirable situations when put into practice, with parties finding that they cannot execute them or that unanticipated circumstances arise which cannot be resolved without resorting to costly and lengthy litigation. If a contract can be both formally specified and *verified*, that is, checked for undesirable pathological features, then not only is its negotiation conducted more effectively but also its subsequent performance is smoother.

A number of formal techniques have proved effective in finding pathological features in specifications of hardware (and software) systems. What follows presents the essential features of such techniques and discusses how they can be applied in the case of contracts[1]. The verification techniques discussed here operate on Petri net representations. Petri nets were chosen because there was available software that implements these verification techniques; in principle though the verification techniques can be extended to any state-based description. In what follows we show how Petri nets are constructed from initial state diagrams for a contractual scenario that we use as an example.

---

[1] A more general version of the discussion in this paper can be found in [16].

## 2. An Overview of the Verification Technique

Initially, a system is described using a number of Petri nets [12] that model it from different perspectives. In the case of hardware design, one Petri net is constructed for each individual component of the system or for each set of different requirements. In the case of contracts, each Petri net represents one of the parties' view of the business exchange. Representing contracts as Petri nets is not novel and has in fact been attempted by Lee and his associates (cf. [1]). In this approach the separate nets are combined into one large Petri net using a composition algebra. The final Petri net is likely to be too large and complicated to have been reliably designed and safely input by hand. The composition and every other process mentioned in this section, is fully automated. The only human inputs to the system are descriptions of the initial Petri nets and some behavioural requirements. More information on the use of this composition algebra in micro-electronic design can be found in [15] and [14].

The composed Petri net is processed in order to construct a model of its possible states. Behavioural requirements for the model are then expressed in a species of temporal logic and the state model is queried. This process is called *model checking* [7, 11]. Temporal logic enables the expression of properties about all future times from initialisation. Examples of the kinds of questions that can be expressed are "Is there a possible future in which a given request will never be answered?" or "Can the contract ever be in a position in which one party must perform two conflicting actions?". Model checking systems applied to hardware design are reasonably efficient and have reported some notable successes, including the discovery (in minutes) of bugs that eluded months of NASA testing [8]. The verification models and processes involved in model checking are explained in the sections that follow in terms of a contractual example.

### 2.1. *Specification of a Sales Contract through State Diagrams*

This section considers a contractual scenario in which a seller interacts with a purchaser[2]. The parties agree that delivery of goods will happen when and if the purchaser can take such delivery. The parties' contract stipulates that the business exchange will operate in two phases. The first phase concerns the transfer of goods from the seller to the purchaser. The second phase concerns the transfer of funds from the purchaser to the seller in payment of the goods that were exchanged.

The parties agree that the first phase of the transaction will operate as follows: The seller can indicate that he has goods available for delivery (GAV). The purchaser can indicate that he requests goods (RFG) and that he has taken delivery of goods (GAC). The required sequence of events for the transfer of goods from the seller to the purchaser is as follows: GAV, RFG and GAC are initially false (denoted as GAV_F, RFG_F, GAC_F), that is, the seller has no goods available for delivery, the purchaser does not request any goods and has not accepted any goods. When the purchaser requests goods (RFG_T) the seller may assert that goods are available (GAV_T) and the request for goods is discharged (RFG_F). When the purchaser has taken the specified quantity of goods GAC becomes true (GAC_T) and only then may the seller assert that there are no more goods available (GAV_F). Finally the purchaser may set GAC back to false and the cycle may repeat.

The second phase of the transaction, where funds are transferred between the purchaser and the seller will operate as follows: The purchaser can indicate that he has funds ready for payment (FAV). The seller can indicate that he requests funds (i.e. that he requests payment) (RFF) and that he has received funds (FAC). The reader may notice that the propositions indicating availability of funds, request for funds and acceptance of funds are of the same form as those indicating availability of goods, request for goods and acceptance of goods. Indeed the sequence of events for the second phase of the transaction is of the same form as the sequence of events for the first phase of the transaction. What follows concentrates therefore on the first phase of the transaction but is applicable to the second phase as well.

The state diagram in Figure 1 illustrates the seller's view of the first phase of the transaction. The seller is initially in state $S_0$ (which corresponds to his being not ready for the transaction, for example because he has no goods available). When the seller is ready (in state $S_1$), if he receives a request for goods from the purchaser he moves to a state where he starts the sale ($S_2$). When the goods have been delivered to the purchaser (that is, from the point of view of the purchaser when he has accepted them, GAC_T) the seller is in state $S_3$, where the sale has finished and he may end the first phase of the transaction.

---

[2] This is a variation of the multi-party trading procedure described in [16], where a single seller interacts with multiple purchasers.

The state diagram in Figure 2 illustrates the purchaser's view of the first phase of the transaction. The purchaser is initially in state $P_0$ (not ready for the transaction, for example because he does not require any goods). When the purchaser is ready and requests goods (state $P_1$) if goods are available (GAV_T) he moves to a state where he starts the purchase ($P_2$). When the goods have been delivered the purchaser is in state $P_3$ where he has finished his purchase and accepted the goods.

The initial specification of our example contract was given in terms of state diagrams[3]. The verification techniques discussed in this section operate on specifications given as Petri nets, which are generalisations of state machines.

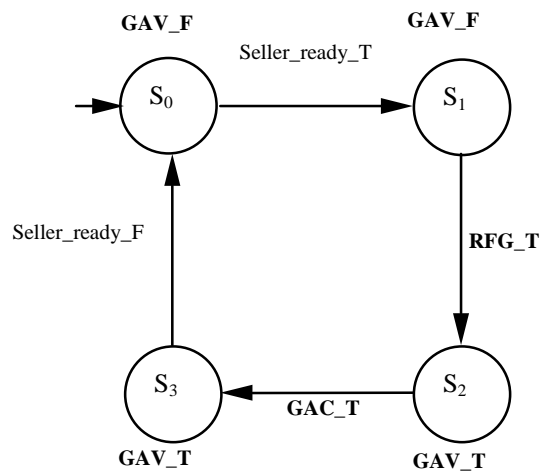

Figure 1 The seller's view of the contract

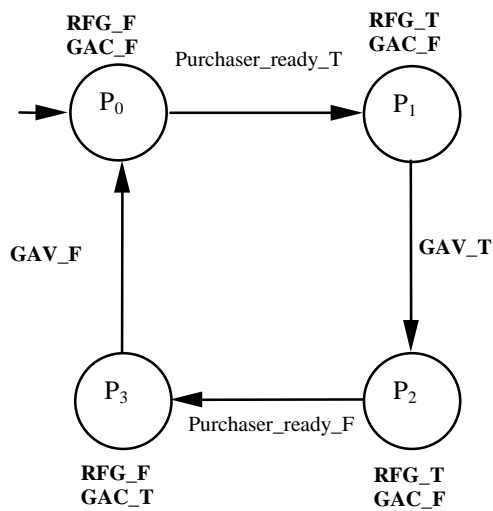

Figure 2 The purchaser's view of the contract

---

[3] See [5] for more details on representing contracts using state diagrams.

## 2.2. *Petri Nets and their Composition*

A Petri net is essentially a bi-partite directed graph, which comprises two sets of nodes: a set of places *P* (represented as circles) and a set of transitions *T* (represented as bars). Arcs connect places to transitions or transitions to places (but not transitions to transitions or places to places). The dynamic behaviour of the system being modelled is represented by tokens (shown as dots) flowing through the net. Each transition in a Petri net has certain requirements (preconditions) and effects (postconditions). At any given time, each place of a net may or may not have a token in it. Places containing tokens are said to be *marked*. A *marking* of a Petri net is a snapshot of the net that indicates what commodities are available at any given time. A transition is said to be *fireable* in a given marking if all of its preconditions are marked. When such a transition fires, its preconditions are unmarked and its postconditions become marked. Dynamic behaviour in a Petri net is obtained as a sequence of markings, each derived by firing one or all of the fireable transitions from the previous marking.

Petri nets are usually given pictorially as in Figure 3, where the single transition has preconditions *p1* and *p2* and postconditions *p1* and *p3*. The transition is fireable when *p1* and *p2* become true (in their respective places) and as a result it renders *p1* and *p3* true.

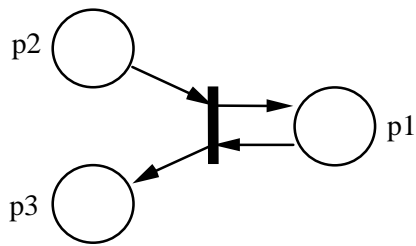

Figure 3 Example Petri Net with three places and one transition

There are various flavours of Petri nets, allowing for example transitions to have outputs [6]. Lee and his associates [1] have introduced another variant, called *Documentary Petri Nets*, for modelling trade procedures. In Documentary Petri Nets some of the places are distinguished (and represented as squares rather than circles) to denote documents—Bons *et al.* find this a useful distinction for their method of composing different views of the same transaction.

In the context of the techniques discussed here Petri nets are composed quite differently, so we restrict the discussion to simple Petri nets. However, it should be noted that all the mathematical results and software alluded to here have variants for several different flavours of Petri nets. Moreover, since Petri nets are a generalisation of finite state machines, the techniques described here are directly applicable to any modelling technique based on finite state machines.

Composition of Petri nets is conducted through labelling them by naming some of the transitions with elements from a labelling set *L*. The labels provide synchronisation information (those transitions in the individual nets without labels are not available for synchronisation). The composed Petri net may contain unlabelled transitions, which cannot be synchronised with any transitions in other component nets. The synchronisation information used in the composition is given by partial functions from an event set *E* to the various labelling sets. A transition *t* from one Petri net synchronises with a transition *t'* from another, if the labels of the two transitions are both mapped to by the same event in the event set *E*. The seller's state diagram that we saw in Figure 1 translates to the Petri net illustrated in Figure 4. The purchaser's state diagram that we saw in Figure 2 corresponds to the Petri net shown in Figure 5.

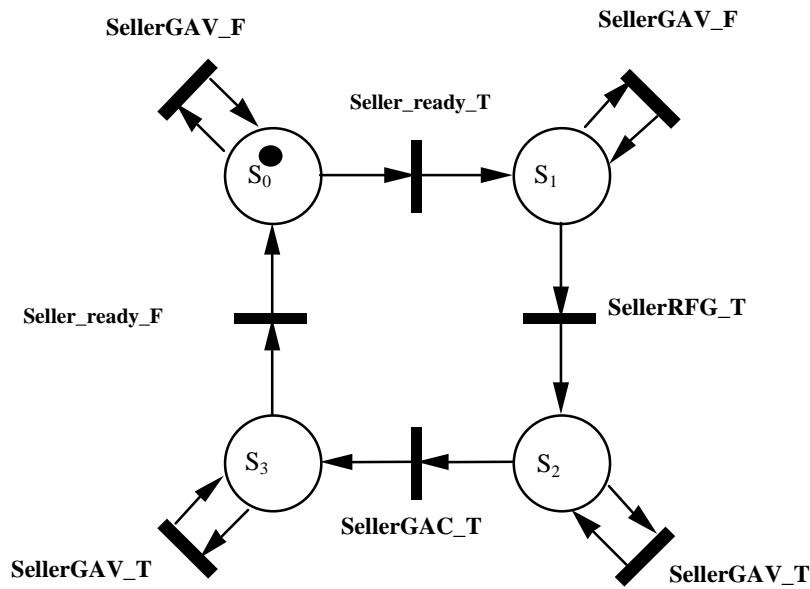

Figure 4 The seller's Petri net of the transaction

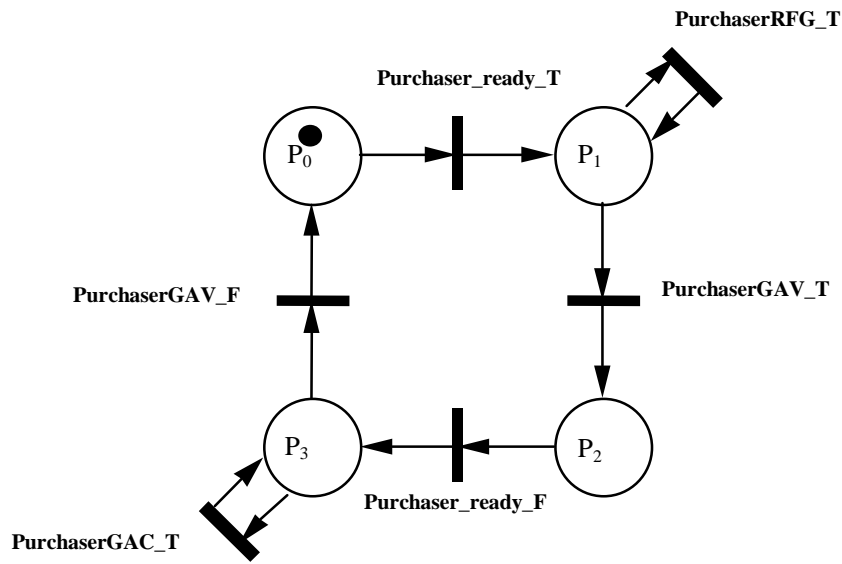

Figure 5 The purchaser's Petri Net of the transaction

The self-loops on some of the places are necessary in synchronisation. The specification of the transaction is completed by adding the synchronisation information, that is, by providing an event set and the information about how each event is viewed in each net.

In this case the event set is:

$E - \{RFG, GAC, P\_GAV\_T, P\_GAV\_F\}$

The events are "request for goods", "goods accepted" and goods becoming available (P_GAV_T) or unavailable (P_GAV_F) to the purchaser. The information about each of the parties' view of these events is given by the following partial functions, which map events (shown on the left of the $\mapsto$ symbol) to transitions in each of the Petri nets that model each party's view of the transaction:

Seller:
$$RFG \mapsto SellerRFG\_T$$
$$GAC \mapsto SellerGAC\_T$$
$$P\_GAV\_T \mapsto SellerGAV\_T$$
$$P\_GAV\_F \mapsto SellerGAV\_F$$

Purchaser:
$$RFG \mapsto PurchaserRFG\_T$$
$$GAC \mapsto PurchaserGAC\_T$$
$$P\_GAV\_T \mapsto PurchaserGAV\_T$$
$$P\_GAV\_F \mapsto PurchaserGAV\_F$$

The composed Petri net with one purchaser and one seller is illustrated in Figure 8. The composition allows us to combine more than two component Petri nets (say, in the case of multi-party transactions) but this example is intended to illustrate how difficult it would be to describe directly the resulting Petri net for more components.

In a composed Petri net there are two kinds of transitions: some are transitions from the component Petri nets, for which there is no synchronisation information. Such transitions are merely carried over from the component nets to the composed net. The other transitions are the result of synchronising one transition from each component Petri net, in accordance to the event set. Consider for instance the two component Petri nets in Figure 6 and suppose that the synchronisation information available is that an event $x$ synchronises transitions $a1$ and $b1$, while there is no synchronisation information for transition $b2$. That is, the event set is $E = \{x\}$ and we have the following partial functions defined for each component Petri net:

A:     $x \mapsto a1$
B:     $x \mapsto b1$

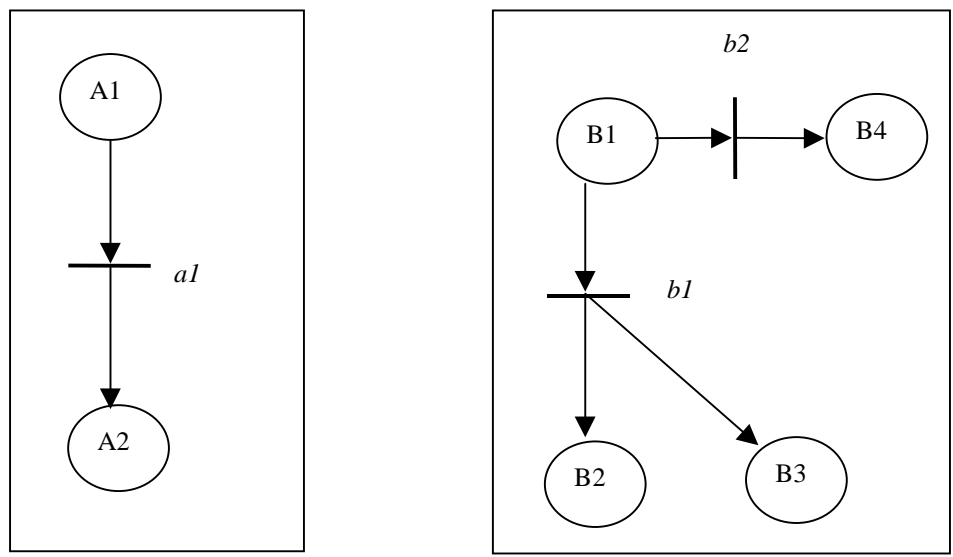

Figure 6 Example Component Petri Nets (A and B)

In their composition, which is shown in Figure 7, the transition marked (*a1, b1)* corresponds to the two synchronised transitions of the corresponding component Petri nets, while the transition *b2* is carried over from Petri net B. A composed transition such as *(a1, b1)* has as input and output places all the input and output places of its component transitions, as can be inferred from the example in Figure 7.

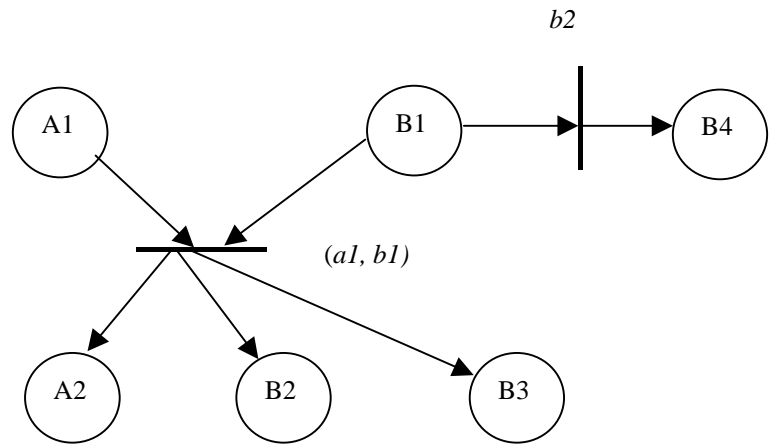

Figure 7 The Composition of Petri Nets A and B

More details on the composition of Petri nets can be found in [14, 15].

## 2.3. State Models

State models (from a systems modelling perspective) are directed graphs or finite state machines without the inputs, that show the possible state changes of a system. State models are generated as the markings of Petri nets. In our example, the initial marking has $S_0$ and $P_0$ marked in the individual Petri nets of seller and purchaser. This leads to the state model shown in Figure 9.

If we had a multi-party scenario, the state model would not have been deterministic. The next stage of verification is to test the truth of various conditions of the state model. The logics used in model checking are all species of propositional temporal logic. The most frequently used is called Computation Tree Logic (or CTL) [3].

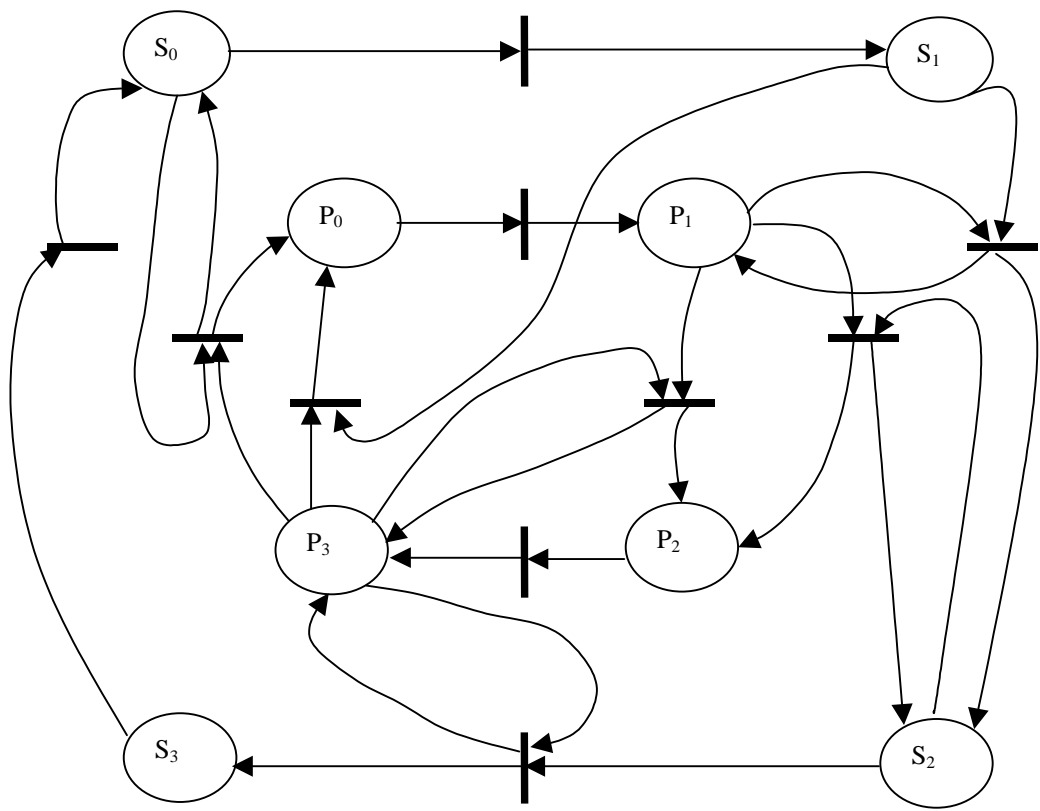

Figure 8 Composed Petri Net of seller and purchaser

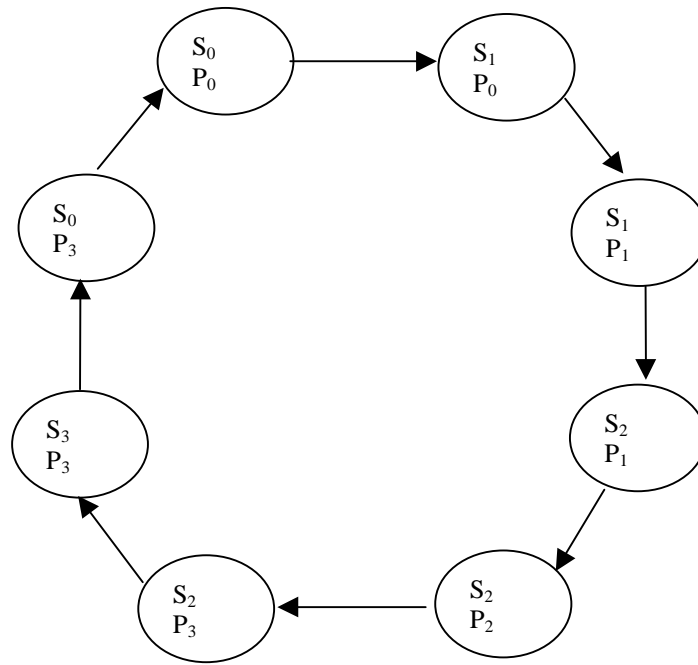

Figure 9 State model of composed Petri net

## 2.4. Computation Tree Logic

Well-formed formulae in CTL are atomic propositional ones or compounds formed by using logical connectives and some temporal operators. Semantic relations are defined between states and formulae. Expressions of the form $s \models \varphi$, where $s$ is a state and $\varphi$ is a formula, mean that $\varphi$ is true in state $s$. The definition of $\models$ is shown in Table 1:

Table 1 Semantics in CTL

| | |
|---|---|
| $s \models \neg \varphi$ | Iff $s$ does not satisfy $\varphi$. |
| $s \models \varphi \vee \psi$ | Iff $s$ satisfies at least one of $\varphi$ and $\psi$ |
| $s \models \varphi \wedge \psi$ | Iff $s$ satisfies both of $\varphi$ and $\psi$ |
| $s \models \varphi \rightarrow \psi$ | Iff $s \models (\neg \varphi \vee \psi)$. |
| $s \models E(\varphi)$ | Iff in some immediate successor state of $s$, $\varphi$ holds. That is, it is possible for $\varphi$ to hold in the next state |
| $s \models A(\varphi)$ | Iff in all immediate successor states of $s$, $\varphi$ holds. That is, it is necessary that $\varphi$ will hold in the next state. |
| $s \models F(\varphi)$ | Iff in some future state reachable from $s$, $\varphi$ holds. That is, $\varphi$ holds eventually. |
| $s \models G(\varphi)$ | Iff in all future states reachable from $s$, $\varphi$ holds. That is, $\varphi$ holds (globally) for ever after $s$. |
| $s \models (\varphi \text{ Au } \psi)$ | On all paths from $s$, $\varphi$ holds at all states until the first state in which $\psi$ holds ($\psi$ is not assumed to ever hold). |
| $s \models (\varphi \text{ Eu } \psi)$ | On some maximal path from $s$, $\varphi$ holds at all states until the first state in which $\psi$ holds ($\psi$ is not assumed to ever hold). A maximal path is either a path that comes to a dead-end or an infinite path. |
| $s \models (\varphi \text{ EU } \psi)$ | On some path from $s$, $\varphi$ holds at all states until the first state in which $\psi$ holds ($\psi$ is assumed to eventually hold). |
| $s \models (\varphi \text{ AU } \psi)$ | On all paths from $s$, $\varphi$ holds at all states until the first state in which $\psi$ holds ($\psi$ must eventually hold on every maximal path). |

The difference between u and U is subtle. Consider the state model of Figure 10, where letters outside states are the names of the states and letters inside the states represent properties that are true in the states.

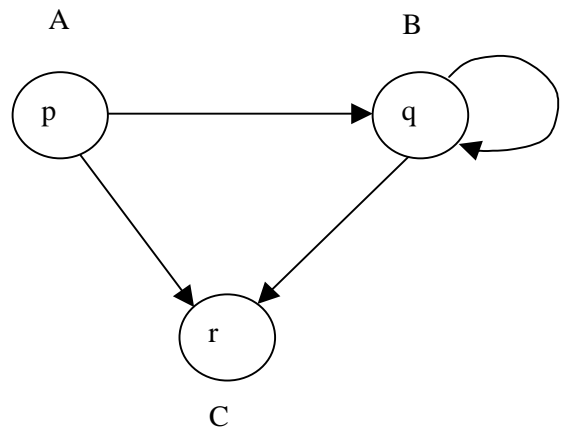

Figure 10 Example state model

There are infinitely many maximal paths from A: there is a length-one path to C; there are paths from A to B staying in B for a number of cycles and then moving to C; and there is an infinite path that goes to B and stays there by repeatedly going round the loop. On all of those paths, q remains true until r is true. That is:

$A \models (q \text{ Au } r)$

On the infinite path r is never true, therefore:

$A \models \neg (q \text{ AU } r)$

There are some paths in which r is eventually true:

$A \models (q \text{ EU } r)$

The distinction between u and U does not show up in our example. However, when more than two parties are admitted (say one more purchaser) the distinction does show up: a purchaser making a request for goods will persist until he is satisfied. However, with two purchasers, if the contract stipulates that the sale starts when all purchasers are ready to take goods, and the second purchaser never requests goods, then the first purchaser's request will never be satisfied.

## 2.5. *Verification*

One condition, a form of liveness that we might wish to check, is that from every state the transaction can eventually return to its initial state. In other words, we wish to establish that the transaction between seller and purchaser will eventually terminate (that the contract between them will at some point be discharged). The condition is expressed below, where B is the initial state:

$B \models G(EF(S_0 \wedge AP_0))$

Some safety conditions we have also checked for the example presented here are:

- The seller should assert GAV_T only when the purchaser is ready (in state $P_1$). In other words, we wish to establish whether there is a situation in which the seller has goods available for delivery but the transaction cannot be realised because the purchaser is not ready to accept delivery:

    $B \models G((\neg SellerGAV\_T \wedge ESellerGAV\_T) \rightarrow P_1)$

- Once GAV is true, it should remain true until the purchaser has accepted goods (that is, the purchaser is in state $P_3$). In other words, we want to establish whether it is possible for the seller to retract goods, once the transaction has started, before they have been accepted by the purchaser:

    $B \models G(SellerGAV\_T \rightarrow (SellerGAV\_T \text{ AU } P_3))$

- The purchaser should not accept goods until GAV is true, that is, the purchaser should not enter $P_2$ until GAV_T. In other words, we want to establish that there is no situation in which the purchaser's obligation to pay for goods is active before such goods are available and delivered to him:

  $B \models G[(\neg P_3 \wedge EP_3) \rightarrow SellerGAV\_T]$

These liveness and safety conditions were checked and all hold for the contract defined in our example.

Checking simple conditions (that involve a state and its immediate successor) is straightforward and computable; the algorithm involves looking ahead at most one step [14]. For each of the more complicated temporal constructs, the model checking algorithm uses iterative search. The method is computable because it can be proven that these iterative techniques all finish after a number of steps bounded by the length of the longest non-looping path in the state space. The model checking algorithm keeps performing iterations until the same result is reached twice (*fixed point*). It can be proven that a fixed point can be reached in bounded time. However, the bound on time is the number of states of the graph, which is at worst exponential on the number of variables. This is called the *state explosion problem*. The example used in this section is relatively small and the number of states was manageable. In the general case, the state model is transformed into another model called a *binary decision diagram* (BDD). The BDD is still exponential in the number of variables but it admits reduction techniques to yield a *reduced ordered binary decision diagram* (ROBDD), which is usually considerably smaller than the original state space[4]. The ROBDD is then queried using a species of temporal logic[5]. More details on the transformation of state spaces into ROBDDs can be found in [11] and a shorter discussion is in [16].

## 3. Discussion

Model checking has proved to be a powerful technique for verifying high-level behaviour of hardware systems. The example that was discussed in this paper showed how the technique can be applied in a contractual setting to check a business transaction (that comes about after two parties have entered a contract) against some temporal conditions. As we saw, existing model checkers operate on two specifications: one, which is the operational specification of system behaviour expressed as Petri nets and the other, which is a declarative specification of behavioural requirements expressed in temporal logic. In a contractual setting, the system being modelled is the agreement. The specification of its behaviour emerges by taking an operational view of the agreement, expressed using Petri nets, as illustrated by the example. Behavioural requirements are requirements about the operational view of the agreement and these can be expressed in temporal logic. By representing agreements in this way, existing formal methods for the verification of hardware and software systems are directly applicable to the verification of business transactions.

For the example discussed in this section bespoke software [14] was used to construct the composed Petri net and to check it. This choice was primarily made because these tools were readily available. Any standard CTL-based model checker would be suitable, such as CheckOff, which is currently marketed by Siemens[6]. The techniques described here were also tried on examples of trade procedures discussed in Lee [10]. Of those, the one that appears to be the most complex concerns a sales contract between a Buyer and a Seller, where a Bank provides a letter of credit and a Carrier is subcontracted to deliver the goods. The abstract view of the transactions between the four parties of the trade procedure is illustrated in Figure 11:

---

[4] The reduction techniques have been used to reason effectively with systems with more than $10^{20}$ states. [2].

[5] In our example we used CTL but other model checkers operate on different temporal logics such as interval temporal logic or linear temporal logic.

[6] In fact CheckOff has been tried on larger specifications for hardware and was faster than the model checker developed by Zimmer & McDonald [14].

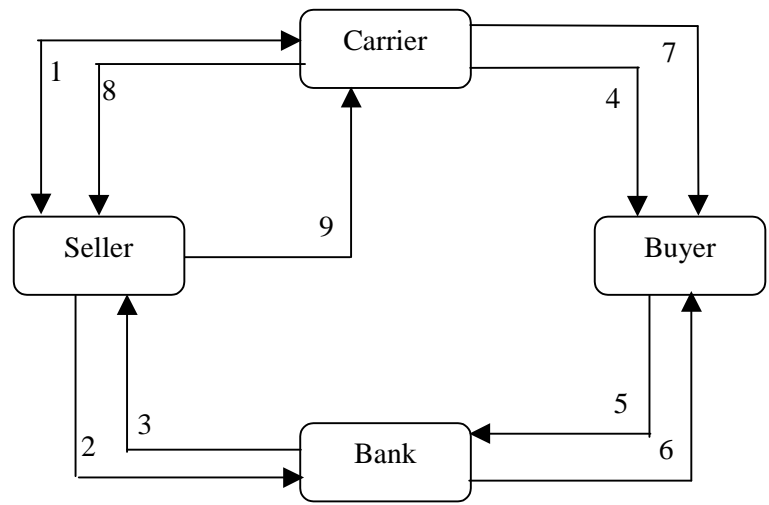

Figure 11 Sales Contract with Carrier Sub-Contracting and Letter of Credit

The various transactions that take place between the parties involved in the trade procedure are as follows:

1. Seller makes goods available to Carrier for dispatch. Carrier provides Seller with certificate that goods have been dispatched.
2. Seller supplies Bank with certificate that goods have been dispatched and certificate that the Buyer owns the goods.
3. Bank pays Seller the required price for goods.
4. Carrier notifies Buyer that goods have arrived.
5. Buyer instructs Bank to pay the required price and interest.
6. Bank sends to Buyer certificate that he owns the goods.
7. Buyer presents Carrier with certificate that he owns the goods, Carrier gives the goods to Buyer, and Buyer gives Carrier receipt.
8. Carrier provides Seller with receipt issued by Buyer.
9. Seller pays Carrier the agreed price for delivery.

The similarity between this seemingly complex example and the simple scenario that we used in this paper should be obvious. In the simple case, we considered only the part of a trade procedure that is relevant to the transfer of goods. In Lee's example, there are goods, funds and various certificates that are exchanged between parties. A Petri net can be constructed showing each of the four participants' views of the aspects of the transaction that are relevant to him and these can be composed and checked against behavioural requirements (Lee [10] lists some such requirements, for instance that transition 2 in the list above must happen before transition 3).

The techniques discussed here were also applied to a few trade procedures that were extracted from sample contracts that concerned the supply of natural gas from hydrocarbon field owners. Our experimentation suggested that the larger the trade procedure that is modelled, the more individual Petri nets need to be formulated so that they can be subsequently composed using the Petri net composition tool. The difficulty in this process lies in establishing the procedure, aspects of which may be described in different parts of a large contractual document, rather than modelling and model checking it. It is also difficult sometimes to establish the precise behavioural requirements that we would want the procedure to satisfy.

We found that the operationalisation of the transaction and its representation as Petri nets conceals the distinction between ideal and actual behaviour. Obligatory, permissible or prohibited actions that parties may perform during the transaction are interpreted and incorporated in the model implicitly, rather than explicitly.

Such interpretation might be incorrect or incomplete (for example a safety or liveness condition might be omitted or misinterpreted). The verification process establishes whether a given operational specification is correct against a declarative one. It does not however address whether *each* of the two specifications is correct or complete.